\documentclass[12pt]{article}
\usepackage{graphicx}
\usepackage{amssymb,amsmath,amsfonts,palatino,amsthm}
\usepackage{enumerate}
\newcommand{\f}{\begin{equation}}
\newcommand{\ff}{\end{equation}}

\newcommand{\beq}{\begin{quote}}
\newcommand{\eeq}{\end{quote}}

\newcommand{\ket}[1]{\vert #1\rangle}

\begin{document}

\title{Counting Steps: A New Approach to Objective Probability in Physics}

\author{Amit Hagar\thanks{Indiana University, Department of History \& Philosophy of Science, Bloomington}$\;$ and  Giuseppe Sergioli\thanks{Universit\'a di Cagliari, Sardegna, Italy}}

\date{\today}

\maketitle
\begin{abstract}

\noindent We propose a new interpretation of objective deterministic chances in statistical physics based on physical computational complexity. This notion applies to a single physical system (be it an experimental set--up in the lab, or a subsystem of the universe), and quantifies (1) the difficulty to realize a physical state given another, (2) the `distance'  (in terms of physical resources) of a physical state from another, and (3) the size of the set of time--complexity functions that are compatible with the physical resources required to reach a physical state from another. This view (a) exorcises ``ignorance'' from statistical physics, and (b) underlies a new  interpretation to non--relativistic quantum mechanics. \end{abstract}

\section{Introduction}
\noindent Probabilistic statements in a deterministic dynamical setting are commonly understood as epistemic (Lewis ,1986). Since in such a setting a complete specification of the state of the system at one time -- together with the dynamics -- uniquely determine the state at later times, the inability to predict an outcome exactly (with probability $1$) is predicated on the notion of ignorance, or incomplete knowledge. Such a subjective interpretation is natural in the context of classical statistical mechanics (SM), where a physical state is represented as a point on phase space, and the dynamics is a trajectory in that space, or in the context of Bohmian mechanics, where the phase space is replaced with a configuration space and the ontology is augmented with the quantum potential, but recently it has been suggested as a viable option also in the context of orthodox non--relativistic quantum mechanics (QM), where the state is represented as a vector in the Hilbert space, and the dynamics is a unitary transformation, i.e., a rotation, in that space (Caves, Fuchs, \& Schack, 2002). In all three cases the dynamics is strictly deterministic, and the only difference -- apart from the representation of the state -- is in the character of the probabilities: in classical SM or Bohmian mechanics they are subsets of phase space (or configuration space) obeying a Boolean structure; in QM they are angles between subspaces in the Hilbert space obeying a non--Boolean structure, whence the famous non--locality, contextuality, and the violation of Bell's inequalities.

Such an epistemic notion of probability in statistical physics appears to many inappropriate. The problem is {\em not} how lack of knowledge can {\em bring about} physical phenomena (Albert, 2000, p. 64); it can't. Neither is it a problem about ontological vagueness (Hagar, 2003). Rather, the problem is that an epistemic interpretation of probability in statistical physics, be it classical SM or QM, turns these theories into a type of statistical inference: while applied to physical systems, these theories become theories about epistemic judgments in the light of incomplete knowledge, and the probabilities therein do not represent or influence the physical situation, but only represent our state of mind (Frigg, 2007; Uffink, 2011). 

In recent years an alternative, objective view of probability, has been defended, both in the foundations of classical SM and in the context of Bohmian mechanics in the foundations of QM, based on the notion of {\em typicality} (Maudlin, 2007). Typicality claims tell us what {\em most} physical states are, or which dynamical evolution are {\em overwhelmingly more likely\/}, by assigning measure $1$ to the set of such states or the set of such dynamical evolutions.\footnote{Examples are ``{\em most} quantum states are mixed, i.e., entangled with their environment'', or ``{\em most} systems relax to thermodynamic equilibrium if left to themselves''. } Such claims make an analytical connection between a deterministic dynamics and a characterization of certain empirical distributions, hence can be interpreted as objective, having nothing to do with one's credence or state of knowledge. With this notion, or so the story goes, one can treat probabilistic statements in a deterministic physical setting as arising from an objective state of affairs, and the theories that give rise to these statements as theories about the physical world, rather than theories about our state of knowledge.

This objective view of probability, however, is not problem--free. First, as its proponents admit (Goldstein {\em et al.} 2010), the notion of typicality is too weak: a theorem saying that a condition is true of the vast majority of systems does not prove anything about a concrete given system. Next, the notion lacks logical closure: a pair of typical states is not necessarily a typical pair of states, which means that ``being typical'' is not an intrinsic property of an initial condition, not even for a single system, but depends on the relation between the state and other possible initial conditions (Pitowsky, 2011). Possible ways around these difficulties have been suggested,\footnote{Maudlin (2007, p. 287), for example, rejects the requirement to assign probabilities to single systems, and Pitowsky (2011) proposes to retain most of the advantages of typicality but to  retreat to a full--fledged Lebesgue measure, with its combinatorial interpretation.} but while these problems may be circumvented, there exists a deeper lacuna underlying the notion of typicality which threatens the entire project.

The point is that typicality claims depend on a specific choice of measure, usually the Lebesgue measure or any other measure absolutely continuous with it. But what justifies this choice of measure when an infinite number of possible measures are equally plausible (Hemmo \& Shenker, 2011a)? Moreover, even if we have established somehow that, relative to a preferred choice of measure, a certain set of states $T$ is typical, i.e., its members are overwhelmingly more probable with respect to all possible states, what justifies the claim that we are likely to observe, or ``pick--up'', members of that set $T$ more often than members of the complement set $\tilde T$? After all, the measure we have imposed on the space of all {\em possible} potential states $(T \cup \tilde T)$ need not dictate the measure we impose on the space of our {\em actual} observations. Indeed, while under the choice of the Haar measure ``most'' quantum states are mixed, hence entangled with their environment, we can still realize (hence observe) pure states in the lab, at least to a certain extent. In what sense, then, are these states ``rare''?

The twofold problem of justification of the measure is, on final account, a manifestation of the problem of induction (Pitowsky, 1985, pp. 234--238). On a strictly empiricist view, the effort to justify typicality claims is just another (futile) attempt to give demonstrations to matters of fact, or to derive contingent conclusions from necessary truths. The point here is that there is no surrogate for experience in the empirical sciences, and that inductive reasoning is the best one can do in one's attempts to understand the world. We tend to agree with the above criticism, but we also believe the alternative (or rather the lack thereof) it leaves us with is equally unsatisfactory. While typicality arguments do seem to achieve too much, the above pessimism seems to leave us with too little: not only are we unable to make sense of objective deterministic chances on this empiricist account, we are also unable to justify the standard statistical methods in the scientific practice. These methods are commonly understood within the context of the frequentist approach to probability, and yet the latter approach is based on typicality arguments (these appear, e.g., implicitly in the weak law of large numbers, or explicitly in the definition of a random sequence), and so the criticism raised against typicality equally undercuts the attempts to apply frequentist methods in the empirical sciences (Hemmo \& Shenker, 2011b). 

Since we believe there is more to statistical physics than statistical inference, we here propose a new non--frequentist interpretation of physical probability as an alternative to typicality. Our notion is objective, dynamical, applies to individual states or systems, and its definition requires no convergence theorems. Die hard empiricist as we are, we offer little relief from the problem of justification of the measure imposed on the space of all possible states.  We do address directly, however, the problem of justification of the measure imposed on the space of all actual states: on our view, objective physical probabilities are transition probabilities that supervene on the time--complexity of the actual dynamical evolution. Our proposal is thus consistent with the above criticism marshaled against typicality and frequentism, and can serve as a viable alternative to the current epistemic view of probability in statistical physics, turning the latter once again into a physical theory about the natural world.

The paper is structured as follows. In section 2 we spell out the basic assumptions behind our proposal. Warming up, in section 3 we show how by equating ``probable'' with ``easy'' (in terms of computational complexity), one can assign measure $1$ ($0$) to a set of states whose realization requires a dynamical evolution with polynomial (exponential) time--complexity. In section 4 we move to a full--fledged definition of objective probability that quantifies how hard it is to realize a physical state, and measures (in terms of physical resources) the `distance' between any such pair of states. In section 5 we apply our new interpretation to classical, statistical, and quantum mechanics. In doing so we introduce a new interpretation to the probabilities of QM.  Section 6 concludes.

\section{Assumptions}

\noindent We start by spelling out the five basic assumptions that underlie our models. They are {\em Determinism\/}, $\mathtt{P}\subset \mathtt{EXPTIME}$, {\em Boundedness}, {\em Discreteness\/}, and {\em Locality\/}. These assumptions are working hypotheses in the framework from which our interpretation of probability stems, namely, physical computational complexity. In this framework (Geroch \& Hartle, 1986; Pitowsky, 1990; Pitowsky, 1996), the performance of physical systems is analyzed with notions and concepts that originate in computational complexity theory, by approximating dynamical evolutions with a discrete set of computational steps to an arbitrary degree of accuracy. These assumptions help us delineate the two probability spaces in our models: the space of physically allowable states, and the space of physically allowable dynamical evolutions.

\subsection*{A. Determinism}
\noindent  Our models rest on the assumption of strict determinism. This assumption follows from the strong physical Church--Turing thesis (PCTT henceforth),\footnote{The {\em physical} Church--Turing thesis is logically independent of the original Church--Turing thesis. See e.g., (Pitowsky \& Shagrir, 2003).} according to which actual dynamical evolutions of physical systems in our world can be regarded as computations carried by deterministic Turing machines. Agreed, some physical theories do allow {\em in principle} for non--Turing--computable phase trajectories (trajectories that cannot be represented by recursive functions), and, in addition, there exists a vast literature on the physical possibility of supertasks and ``hypercomputation'', that aims to show that Turing--computability is not a natural property, and need not apply {\em a priori} in the physical world. Nevertheless, if the strong PCTT holds, then  {\em as a contingent matter of fact\/}, non--Turing--computable trajectories are ruled out, and  the above, rather contrived, counterexamples are not realizable in the actual universe.\footnote{So far there are two such counterexamples: Pour el \& Richards' (1989) wave equation in 3 dimensions and Pitowsky's (1990) spacetime model that allows finite--time execution of an infinite number of computational steps. See also (Hogarth, 1994) for an elaboration on the latter, and (Earman \& Norton, 1993) for further discussion.} In what follows we thus disregard naked singularities, closed timelike curves, non--globally hyperbolic spacetime models, ill--posed problems, divergences, and the like, adhering to the idea that every dynamical evolution takes a physical state to one and only one physical state.\footnote{Note that from a strictly dynamical perspective, quantum {\em dynamics} is fully deterministic: Schr\"odinger's equations takes any quantum state to one and only one quantum state.}

\subsection*{B. $ \mathtt{P} \subset  \mathtt{EXPTIME}$}
\noindent The fact that each computation requires physical resources (energy and time) that increase with the size (the number of degrees of freedom) of the system allows us to classify different dynamical evolutions as either ``easy'' (i.e., having polynomial time--complexity such as $O(n^c)$) or ``hard'' (i.e., having exponential time--complexity such as $O(c^n)$).\footnote{Here $n$ is the input size -- in our case the dimension of the system at hand, and $c$ is a (rational, as we shall assume below) coefficient.} That there exists a meaningful difference between these two classes (and between different degrees of time--complexity within each class) is the consequence of the Time Hierarchy Theorem (Hartmanis \& Stearns, 1965).  

\subsection*{C. Boundedness}
\noindent  Assumption (A) allows us to apply the machinery of complexity theory to dynamical evolutions, by treating them as computations.  Assumption (B) allows us to classify states (and the dynamical evolutions that realize them) as ``easy'' or ``hard''. Assumption (C) allows us to impose upper and lower bounds on the set of all possible dynamical evolutions in the actual universe, based on the following two facts:
\begin{itemize}
\item The physical resources (energy and number of particles) in the universe are bounded from above; beyond a certain degree of an exponential or a polynomial time evolution, the next computational step would require resources that would supersede this bound.
\item  The minimum number of computational steps is 1, and so for a given $n$ (the size of the system) there always exists a lower bound on the set of possible dynamical evolutions below which the number of steps required for this input size $n$ is smaller than 1.
\end{itemize}

\subsection*{D. Discreteness}
\noindent Assumption (D) allows us to discretize the set of the physically allowable dynamical evolutions. Two facts warrant the elimination of real coefficients in our classification of dynamical evolutions into time--complexity classes. First,  each dynamical evolution is governed by a Hamiltonian (the total energy function). Second, the time--energy uncertainty relation limits the ability to resolve arbitrary energy differences between any two Hamiltonians (Childs, Preskill \& Renes, 2000; Aharonov, Massar, \& Popescu, 2002). This means that we cannot distinguish between two unknown Hamiltonians with infinite precision, hence the space of possible Hamiltonians is discrete.

\subsection*{E. Locality}
\noindent Finally, and consistent with the current state of affairs in physics,
in physically realizing the Hamiltonians that govern the dynamical evolutions, we allow only local interactions. 

\subsection*{Models}
\noindent The above assumptions allow us to propose two possible models for objective physical probability. We do not claim that these models are unique, optimal, or in any sense canonical. Our purpose is only to demonstrate that it is  possible to define a finite notion of objective probability in physics on the basis of considerations from physical computational complexity.

Our first model is constructed on the space of all possible physical states of a given physical system with a given number of degrees of freedom $n$ in a given moment in time $t$, confined to a given energy shell $E$. Each such state, given assumption (A), is a result of a certain dynamical evolution, which, in turn, is generated by a certain Hamiltonian. If we assume further that all dynamical evolutions start from a common, ``mother'' state, say, the initial state of the universe,  we can then assign (using assumptions (B)--(E)) a {\em non\/}--uniform probability distribution on the set of all possible states according to the time--complexity of the dynamical evolution that realizes each state.\footnote{The move from the space of states to the space of dynamical evolutions is licit given our assumptions (A) and (B) above: in any given moment in time, and  for each physical state, there can be one and only one time--complexity class of dynamical evolutions that ``realizes'' it from the common ``mother'' state. This means that while many time--complexity classes of dynamical evolutions can realize the same state, no two evolutions that belong to different time--complexity classes can do so {\em at the same moment} if they start at the same common ``mother'' state.}  As we shall see below, for a large number of degrees of freedom, this assignment has interesting consequences. 

Our second model is constructed on the space of all possible dynamical evolutions. Here, again, to precisely define the notion of objective physical probability we require the above triplet, i.e., the number of degrees of freedom $n$, time $t$, and energy $E$. Given such a triplet, we construct a probability space out of a functional that relates the power ($E/t$) of a computation -- seen as a dynamical evolution from one state to another -- with the relative size of the set of the possible dynamical evolutions that are compatible with it. Our probability function is thus a distance measure on the above functional, that quantifies how hard it is to realize a state, or how far a given system is from that state, in terms of the physical resources available to it, relative to the required resources.

\section{Warming Up: Not All States Are Born Equal}
\noindent The standard story about typicality (in classical SM, or in Bohmian mechanics in the context of QM) requires the notion of equiprobability, or a uniform measure, relative to which a set is declared {\em typical\/}. The foundations of SM are saturated with failed attempts to justify this choice of measure,\footnote{See, e.g., (Sklar 1993, pp. 156--195) for a summary.} the most famous of which is Boltzmann's ergodic hypothesis.\footnote{That the ergodic hypothesis falls short of justifying the assumption of equiprobability follows from three facts: (1) many thermodynamic systems are not ergodic, (2) ergodicity holds only at infinite time scales, and (3) such a justification is plainly circular, as it is valid only for a set of ``normal'' states whose measure $1$ is fixed, again, relative to the choice of measure we are trying to justify from the outset. The last two facts are, essentially, equivalent to the recent criticism against typicality.} Our first stab at the notion of physical probability based on computational complexity suggests how to deflate this problem. 

At the crux of the matter lie the primitive notions of {\em number} and {\em counting\/}. The choice of the Lebesgue measure is deemed ``natural'' when one extends the standard notion of counting from the finite case to the infinite one (Pitowsky, 2011). But why treat each state as equal (in number) to another even in the finite case? Agreed, there are physical situations involving symmetries, such as the case of a fair die, that warrant such a treatment (Strevens, 1998), but {\em in general} there is no {\em a priori} reason to count this way. Consequently, in what follows we shall treat physical states in an utterly politically incorrect manner, assigning states with {\em non\/}--equal weights inverse proportionally to the degree of time--complexity of the dynamical evolution they are associated with. As it turns out, few {\em empirical} facts allow us to get as close to equiprobability as one can get in a finite setting, {\em without} relying on any {\em a priori} notion of equiprobability or uniform probability distribution from the outset.
    
Our probability triplet thus consists of the following elements:
\begin{itemize}
\item $\Omega$ is the state space of all physical states of a system with $n$ degrees of freedom on a given energy shell in a given moment in time, obeying the current laws of physics, i.e., realized by a certain physical process whose time--complexity is either polynomial or exponential (``easy" or ``hard"). Note that $\Omega$ is a bounded and discrete set of polynomial functions $\in O(n^c)$ and exponential functions $\in O(c^n)$.
\item $F$ is the $\sigma$--algebra of $\Omega$, i.e., a non--empty class of subsets of $\Omega$, containing $\Omega$ itself, the empty set, and closed under the formation of complements, finite unions, and finite intersections (i.e., $F$ is a discrete, bounded subset of the power set of $\Omega$). The elements of $F$ are possible physical states, realized by physical processes with a combined time--complexity, either exponential or polynomial.
\item $P$ is the probability measure that maps members of $F$ onto $[0,1]$, where $P(\emptyset) = 0$, $P(F) = 1$, such that $P$ is additive.
\end{itemize}

We now proceed to the assignment of measures on sets of states. Let's denote with $\mathtt{Exp}$ and $\mathtt{Poly}$ a partition of $F$ into two subsets with some prior measures $\mu_e$ and $\mu_p=1-\mu_e$.

Next, for every function $f \in (\mathtt{Poly} \cup \mathtt{Exp})$ we define the {\em weight} $\xi$ of $f$ at an arbitrary (natural) point $n$ as:\footnote{Metaphorically, dynamical evolutions in $\mathtt{Exp}$ ``lose weight'' in direct proportion to their degree of time--complexity. This total lost weight is now (non--uniformly) distributed on $\mathtt{Poly}$ in such a way that the dynamical evolutions in $\mathtt{Poly}$ gain relative weights inverse proportionally to their degree of time--complexity.}

\begin{equation}
\xi_{f(n)}=\mu_{e}-\frac{\arctan f^\prime (n)}{\Pi/2}\mu_e 
\end{equation}
when $f(n) \in \mathtt{Exp}$, and
\begin{equation}
\xi_{f(n)}=(1-\mu_e) + [(1-\mu_{e})-\frac{\arctan f^\prime (n)}{\Pi/2}(1-\mu_e)]  
\end{equation}
when $f(n) \in \mathtt{Poly}$.

For every function $f \in (\mathtt{Poly} \cup \mathtt{Exp})$ we define the \emph{probability} of $f$ in an arbitrary (natural) point $n$ as
\begin{equation}
P(f(n)) = \alpha \xi_{f(n)}.
\end{equation}
where $\alpha$ is a normalization parameter given by: 
\begin{equation}
\alpha=\frac{1}{\sum_{i} \xi_{f_i(n)}}.
\end{equation}
It is easy to show that $P(f(n))\in [0,1]$, that $P(\emptyset) = 0$, and that by construction, the total probability is equal to 1. Furthermore, if we take into account the complexity degree of each element of the bounded set $(\mathtt{Poly} \cup \mathtt{Exp})$, we can obtain a partition of $(\mathtt{Poly} \cup \mathtt{Exp})$ where every element of the partition (indicate by $I_1, ... ,I_n$) corresponds to a different degree of complexity. We now define $P(I_1)=\sum_{f_i\in I_1}P(f_i)$. It follows that: \begin{equation}
\forall i,j, P(I_i \cup I_j)=\sum_{f_i \in I_i} P(f_i)+ \sum_{f_j \in I_j}P(f_j).
\end{equation} 
Thus $P$ satisfies Kolmogorov's axioms, and as such it is an admissible probability function.

Note that $\xi$ is inverse proportional to the first derivative of the time--complexity function of the dynamical evolution that realizes the state. Consequently, for a large input size (i.e., a large number of degrees of freedom), the following results hold (see appendix A for details):
\begin{enumerate}[I.]
\item  The set of states whose time--complexity is exponential gets assigned a measure close to zero, while the set of states whose time--complexity is polynomial gets assigned a measure close to one.
\item The polynomial states are (almost) uniformly distributed, i.e., their distribution is (almost) a resolution of the identity.  
\end{enumerate}

We would like to clarify the following two points: 
\begin{itemize}
\item Since the $\sigma$--algebra $F$ is a bounded and discrete, $\alpha$ is always finite (albeit very small for a large $n$). For this reason, in a finite model such as ours, the assignment of actual measure $0$ (or $1$ for that matter), as well as the assignment of equiprobability (as a resolution of the identity on $\mathtt{Poly}$) are fictions, as the set $\mathtt{Exp}$ remains with a finite (albeit very small) measure, and the partition on the set $\mathtt{Poly}$ is never strictly uniform. For a macroscopic system, however, `very small'  is an understatement, and the above partition is very close to uniform. When the state at hand is of the universe as a whole (where $n \approx 10^{80}$), the understatement is literally a cosmic one, and equiprobability of all polynomial time--complexity functions is a practical certainty. 
\item The normalization factor,  $\alpha$, is inverse proportional to the number of possible time--complexity functions, and yet for combinatorial reasons, this number, $i\,$, is directly related to $n\,$, the number of degrees of freedom, hence, in effect, $\alpha$ is inverse proportional to $n\,$. One can object here that we implicitly ``sneak in'' an assumption about equal weights by treating each degree of freedom on equal footing.\footnote{Recall that each time--complexity function represents a specific interaction Hamiltonian that generates it. For the normalization factor to be inverse proportional to the size of the system, i.e., $\alpha \propto n^{-1}$, we must assume that a two--body system possesses {\em less} possible interactions (hence {\em less} possible dynamical evolutions) than a many--body system. This assumption, we argue, follows from the the locality condition.} In response, we stress that the this result rests not on an {\em a priori} notion of counting, but rather on our assumption (E) above and on the {\em empirical} fact that a concatenation of nearest--neighbor interactions is {\em not} physically equivalent to one nonlocal interaction.\footnote{For an analysis of the complexity costs involved in simulating a nonlocal operator with local ones see, e.g., (Vidal \& Cirac, 2002).} In other words, it is because of this (contingent) nature of physical interactions (and the resources they require) that results (I) and (II) above hold.
\end{itemize}

Are these results sufficient to support the common lore, according  to which  ``most''  states are thermodynamic normal, i.e., {\em typical\/}, hence more likely to be observed? The answer is clearly negative, but the reasons for the insufficiency are quite subtle. 

First, note that we have deliberately distributed weights on different time--complexity functions in such a way that favors members of $\mathtt{Poly}$ and disfavors members of $\mathtt{Exp}$ (see equations (1) \& (2) above), but nothing justifies such a distribution -- we could just as well have constructed a symmetric model in which $\mathtt{Exp}$ turned out to be assigned a measure close to $1$ and $\mathtt{Poly}$ a measure close to $0$. This is exactly the problem typicality arguments face (Hemmo \& Shenker 2011a), and in this respect, our model fares no better. What are model {\em does} show, however, is that for a large dimension, and given several plausible assumptions such as ours, equiprobability (or some distribution close to it) holds among members of {\em one of the two sets\/}. It is still a contingent matter of fact {\em which} set   ``wins over'', and in this sense, experience remains the only source for justifying the choice of measure.  

In order to highlight another interesting feature of our model, let's assume that (I) and (II) above hold.  One could still argue that in order to support the common lore, it is necessary to demonstrate that the time--complexity of the dynamical evolutions associated with those normal states is polynomial. According to our model this would endow such states with high probability. The problem is that the common lore also associates thermodynamically normal behavior with non--integrable systems, yet our probability model is discrete. As such, it harbors only integrable, or periodic dynamical systems. One could still observe chaotic behavior in this context,\footnote{There is no compelling reason to associate chaos only with the cardinality of the reals. See (Winnie, 1992) on the idea of ``computable chaos''.}  but this would require redefining notions such as  ``sensitivity to initial conditions'' or ``dynamical instability'' to fit the discrete background, and would also require a careful analysis of time scales.\footnote{At this point, and for the record, let us acknowledge the discrepancy between our discrete model and the continuous nature of the time evolutions. The former is used in computer science; the latter in physics. Both are consistent, and the question whether the former approximates the latter or vice versa, i.e., which is more fundamental, seems to us, at least at the current stage of physics, purely metaphysical.}

But  even such a demonstration would still fall short of supporting the common lore. The standard notion of probability concerns a sequence of events, and in particular, a random choice of such sequences. We could, of course, translate this notion to fit our new definition by exchanging events with physical states, yet nothing constrains us to treat the above probability space of all {\em possible} states as isomorphic to the probability space that contains our {\em actual} observations. In particular, even if on phase space thermodynamic normal states were members of a set of measure $1$ and thermodynamic abnormal states were members of a set of measure $0$, the measure imposed on the space of our actual observations could be {\em different\/}: we could, for example, chose between thermodynamic normal or abnormal states by tossing a fair coin, thus endowing them with equiprobability!

In fact, we know from experience that thermodynamic abnormal states can be realized in the lab. The most famous examples for these are the spin echo experiment (Hahn, 1950) and the Fermi--Pasta--Ulam (1955) discovery of a violation of the equipartition theorem. If we call such `anomalies' ``rare'', we must explain how is it that we can repeat such ``rare'' events ever so often.

In what follows we shall demonstrate how our new view on physical probability can meet these challenges.

\section{One Step at a Time}
\noindent  For an empiricist, the problem of justification of the choice of measure is just a pseudo--problem, experience being the only source of justification required. Our first model, however, may serve as a consistency proof, demonstrating how equiprobability may arise from certain {\em contingent} assumptions about (local) physical interactions and complexity considerations. It thus motivates us to propose a new interpretation of objective probability based on computational time--complexity. We emphasize again that all we offer here is a new {\em meaning} for the term ``probability''. Quantitatively, at this stage we can only propose the {\em conjecture} that, if worked out to its finest details, our proposal will converge to the actual (and so far well--confirmed) probabilities that are currently being employed in statistical physics. We shall say more on this in section (\ref{Ignorance}).  

We thus suggest to interpret objective probability as a physical magnitude that quantifies how hard it is to realize a physical state, given a triplet of physical resources (energy, time, space). Equivalently, this magnitude quantifies how `far' a given physical system is from a certain state in terms of the physical resources available to it, relative to those required for that state's realization. 

\begin{itemize}
\item Take any physical system with dimension $n$ in a given energy state $E$ and in a given moment in time $t$, and let $\Omega$ be the bounded and discrete set of possible dynamical evolutions obeying the current laws of physics, whose time--complexity is either polynomial or exponential (``easy" or ``hard"), that may govern the system's behavior. The set $\Omega$ contains all possible {\em dynamical evolutions\/} that can realize a single actual state.  
\item Given a certain couple $(n, \frac{E}{t})$, where $n$ is the dimension of the state, $E$ is the total possible energy, and $t$ is the total possible time, we consider the set
\begin{equation}
S_{\bar n} = \{ g_n \in \Omega | O(g_{\bar n}) \leq \#(\frac{E}{t})\} 
\end{equation}

where $g_{\bar n}$ is a dynamical evolution that for a given $n$ ``consumes'' at most the resources $E$ in time $t$ (we denote the power allowed for the computation as $\mathtt{Pw}=E/t$).\footnote{\label{consume}By ``consumes'' we mean the following. Take an arbitrary computation. Each computational step ``costs'' the same amount of time; but if, as in our case, the {\em total} time allowed for the computation is fixed, the difference in time--complexity is cashed out in terms of the difference in the frequency of the computation, i.e., the time--difference between any two computational steps. Thus, for a given $n$ and for a given $t$, the higher the degree of time--complexity of the function, the higher the frequency of the computation. Since higher frequency means higher power, by setting a bound on $\mathtt{Pw}$, one immediately sets a bound on the number of computational steps allowed for the computation ($\#$), and subsequently, a bound on the number of time--complexity functions that can realize the computation.}
\item $F$ is the $\sigma$--algebra of $S$, i.e., a non--empty class of subsets of $S$, containing $S$ itself, the empty set, and closed under the formation of complements, finite unions, and finite intersections. The elements of $F$ are dynamical evolutions with a combined time--complexity, either exponential or polynomial. $F$ is thus a subset of the power set of $S$, and is bounded and discrete.
\end{itemize}

Our probability measure $P$ is given by the mapping:
\begin{equation}
\forall A \in F : P_{\{n_A, \left(\frac{E}{t}\right)_A\}}(A) = \frac{|A|}{|S|}
\end{equation}

Where $\left(\frac{E}{t}\right)_A$ is the available power.\footnote{Mathematically speaking, the mapping between $\#$ and $\mathtt{Pw}$ is discontinuous. We can still define $P$ with an integral, however, using an approximation, by embedding this mapping into the continuous function $f$, This embedding has one advantage, namely, it allows us to constrain our model: the curve $\# = f_{\bar n}(\mathtt{Pw})$ relates time--complexity functions (indirectly via the number of steps they require for the computation) with the power of the computation, and is of the general concave form $(n^\alpha\mathtt{Pw})^{1\beta}$ --
since all time--complexity functions are monotonic, have a common origin, and are otherwise non--intersecting, $\forall \mathtt{Pw}_i, \mathtt{Pw}_j , x \;\textit{such that}\;\mathtt{Pw}_i<\mathtt{Pw}_j, \;x \in \mathbb{N}:\; f_{\bar n}(\mathtt{Pw}_{i+x}) - f_{\bar n}(\mathtt{Pw}_i) > f_{\bar n}(\mathtt{Pw}_{j+x}) - f_{\bar n}(\mathtt{Pw}_j)$ -- where $\alpha$ and $\beta$ are free parameters,  constrained by the theorems of probability theory (e.g., independence, conditional probability). See appendix B--E.} 
 To calculate this magnitude we embed it in a continuous function of the general concave form $\#=\left(n^\alpha\mathtt{Pw}\right)^{1/\beta}$, where $\alpha$ and $\beta$ are free parameters.

\begin{figure}
\centering
\includegraphics[scale=0.35]{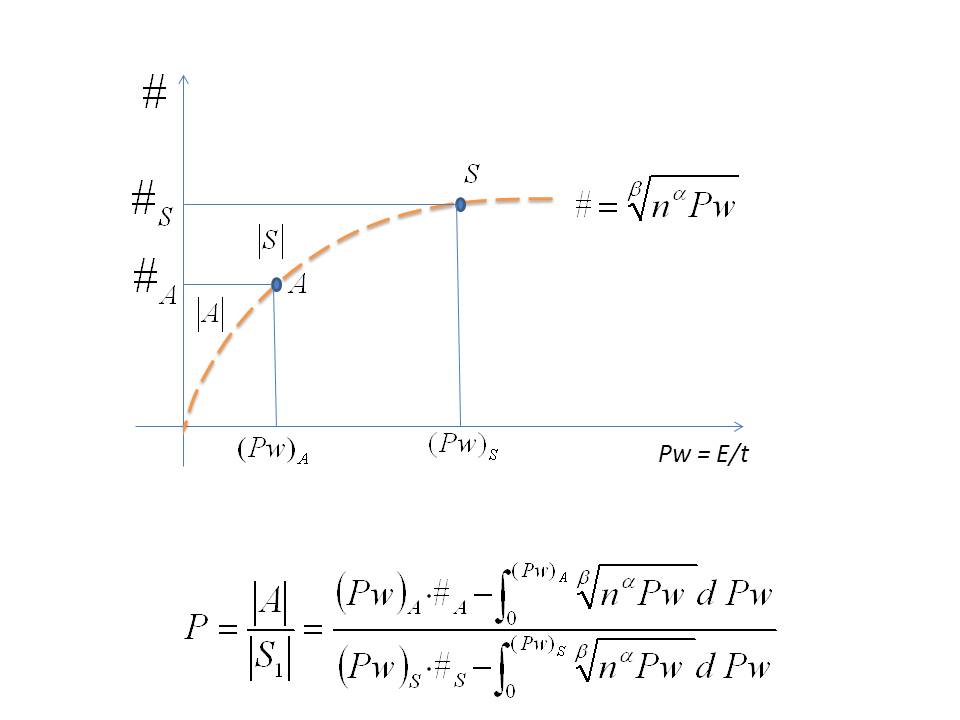}
\caption{Probability from time--complexity}
\end{figure}

By construction $P(A)\in [0,1]$, $P(\emptyset) = 0$, $P(S) = 1$, and $P$ is additive: $\forall A,B \in F$ such that $A\cap B = \emptyset$ , $P(A\cup B) = P(A)+ P(B)$. In order to satisfy further constraints imposed by the axioms of probability theory, the curvature of $\left(n^\alpha\mathtt{Pw}\right)^{1/\beta}$ -- controlled by  $\alpha$ and $\beta$ -- must satisfy further conditions. We spell out these in the appendix.

From a computational complexity perspective, the meaning of $P$ is straightforward:  
\begin{itemize}
\item $P = 0$ means that the desired state is non--Turing--computable, i.e., its realization requires a non--algorithmic process, such as a measurement with infinite precision. 
\item   $P = 1$ means that we are `at' the desired state hence its realization requires constant resources. In complexity theory, such a process would be assigned complexity $O(1)$.
\item $0<P<1$ means that the given system is ``P--distant' from the desired final state. In other words, $P$ denotes the transition probability between the intermediate states, captured by the size of the set of possible time--complexity functions that are compatible with the resources required for that transition.
\end{itemize}

\section{Ignorance of what?}\label{Ignorance}
\noindent Consistent with our goal to turn epistemic probability in statistical physics into an objective one, the notion of physical probability here proposed has nothing to do with one's credence or degrees of belief. It measures, as we have seen, three equivalent {\em physical} properties that each pair of physical states objectively possess:
\begin{itemize}
\item The difficulty (in terms of physical resources) to realize the transition from one state to another. The more probable a state, the easier it is to reach it from a given state with a given amount of resources.
\item The distance (in terms of physical resources) between one state and another. The more probable a state, the shorter is its distance (in terms of physical resources) from the initial state.
\item The relative size of the set of time--complexity functions that are compatible with the above two properties. The more probable a state, the larger is this size.
\end{itemize}
To see how this notion of probability can turn subjective ``ignorance'' in statistical physics into an objective feature of the world, we propose the following intuition. 

\subsection{Classical mechanics}
\noindent Start with classical mechanics. Here the common lore traces probabilistic statements to dynamical sensitivity of initial conditions. Omniscient beings such as Laplace's demon, or so the story goes, could predict with certainty any possible outcome of a dynamical evolution.\footnote{Given our assumption (A), such evolutions are restricted to Turing--computable ones. See (Pitowsky, 1996).} Short of this a power, finite creatures such as ourselves are constrained to introduce error into their predictions. What we suggest here is that this error has physical meaning, captured by our notion of probability.\footnote{On another interesting relation between error and complexity costs see (Traub \& Werschultz, 1999).}

Suppose we would like to predict a certain outcome of an experiment in the lab, described by classical mechanics. Unless we already possess the initial state of our experiment, we need to create, or prepare it. The preparation of the initial state, starting from a specific state we {\em do possess}, requires physical resources. If these are limited, then our probability $P_{\mathtt{complexity}}$ quantifies how far we are from the ideal initial state, or equivalently, what is the error, $\epsilon$, in our preparation, where $P_{\mathtt{complexity}} = 1 - \epsilon$. Thus if resources are insufficient, we start an experiment not in the ideal initial state, but in another, actual state, $\epsilon$--distant from the ideal state, and so, even with deterministic dynamics, we have an error in the final state, which turns out to be different than the one we'd expected. This actual error in the preparation of the state doesn't mean that the system possesses no definite state. On the contrary, the system is always in a definite state; it is just in a state distant from the desired state with a certain error $\epsilon$, which in turn is (inverse) proportional to the resources we employ in the preparation. As our second model shows, this distance can be regarded as a probability measure.

When we move from mechanics to statistical mechanics, we introduce a distinction between micro--states and macro--states. The evolution of the former on phase space is constrained by Liouville's theorem, that tell us that a region of phase space (call it "a blob"), occupied by a set of micro--states all compatible with a certain macro--state, may change is shape but not its volume. The "evolution" of the latter is dictated by the kind of measurements we make, i.e., by the different partitions we impose on phase space. These two evolutions are {\em independent\/},\footnote{see Hemmo and Shenker (2012) for the trouble one gets into when one ignores this independence.} and they allow us to define the transition probability of a physical system from one macro--state to another as the partial overlap between the blobs and the macro--states:
\begin{equation}\label{ProbRule}
P\left([M_1]_{t_1} | [M_0]_{t_0}\right) = \mu\left(B_{t_1} \cap [M_1]\right)
\end{equation}
This means that the probability that a system that starts at a macro--state $[M_0]$ at time $t_0$ (when the size of the dynamical blob $B$ completely saturates the volume $[M_0]$) will end in a macro--state $[M_1]$ at time $t_2$, is given by the partial overlap (the relative size) $\mu$ of the dynamical blob $B$ at $t_2$  with the macro--state $[M_1]$. Note that there is nothing subjective in this kind of transition probability. "Ignorance" here simply means lack of resolution power, i.e., lack of precision or lack of control, which is expressed by the relation between dynamical blobs and macro--states, both of which are objective features of the physical world. 

One can describe the evolution of a dynamical system either by following its dynamical blob, or, equivalently by following the macro--states to which the exact state belongs. In the first description probability signifies lack of precision; in the second, lack of control. We have already shown that our probability measure describes the amount of missing resources for an exact description in the first case. In the second case, we can define our probability as an objective physical magnitude, a transition probability between two macro--states $M_0$ and $M_1$, that signifies how "far" is $M_1$ from $M_0$, where the "distance" $P(M_0,M_1)$  is defined in terms of the physical resources (energy, space, and time) that an observer who observes $M_0$ {\em has\/}, relative to what she {\em needs} in order to observe $M_1$. 

In this sense, probability is an objective measure of the {\em difficulty} to produce the macro--state $M_1$ from the macro--state  $M_0$ given the physical resources (energy, space, and time) at one's disposal. Moreover, this measure is  {\em identical\/}, conceptually and formally, to the one  used in the foundations of statistical mechanics (\ref{ProbRule}), as one can interpret any probability less than $1$ as signifying the lack of physical resources that can allow one to partition phase space into a macro--state more accurately in such a way that it will include {\em all} of the dynamical blob. 

Here is how: when the system starts in a given initial macro--state $M_0$, its  dynamical blob completely saturates the volume $[M_0]$; this is what we mean when we say that we {\em know} the system to be in state $M_0$.\footnote{The opposite case, when we are uncertain of the initial state, can also be accommodated in this framework, by identifying the the error $\epsilon$ with lack of physical resources, and by defining the probability $P = 1 - \epsilon$. In this case $P$ measures the distance between the state we {\em think} the system is in and the {\em actual} state it is in, i.e., the macro--state that the dynamical blob does saturate. In other words, in this case $M_0$ and $M_1$ switch places.}  We now make a measurement, hoping to find the system in $M_1$ after it. In other words, we now carve up phase space into a different macro--state. If we have enough resources available, we can accurately carve the macro--state $M_1$, so that the dynamical blob will, again, saturate its volume $[M_1]$; if we do not have enough resources, only part of the dynamical blob would overlap with $[M_1]$. The empirical conjecture we make, over and above the requirement for conformity with the observed relative frequencies, is that this relative volume of the dynamical blob in $[M_1]$ should be a function of the physical resources we have relative to what we need in order to observe $M_1$ with certainty. This conjecture is in principle testable in many scenarios within control theory, where one is trying to steer a physical process to a desirable outcome. 

\subsection{Quantum mechanics}
\noindent In quantum mechanics the situation is no different. Here the (deterministic!) evolution of a physical system is given by the propagator $U_\beta=e^{-iHt}$ where $H$ is the Hamiltonian of the system. Now, our attempt to estimate this propagator introduces an error, and results in an approximate propagator, $U_\alpha$, where the error is given by  the distance between the two propagators:
\begin{equation}\label{QM1}
d(U_\beta,U_\alpha) = || U_\beta - U_\alpha ||_{op} = \sup_{||\ket{\psi}||=1}  \left| \left| \left( U_\beta - U_\alpha \right) \ket{\psi} \right|\right|_{\mathbb C^2} .
\end{equation}
where $U_\alpha$ and $U_\beta$ are two different rotation operators along, say, the $y$ axis, and $0<\alpha<\beta<\frac{\pi}{2}$. Here, again, the error $\epsilon\,$, and therefore our notion of probability, $P_{\mathtt{complexity}}=1-\epsilon\;$, quantify the {\em distance} (in terms of energy and time) between two physical states -- in this case, the distance, relative to a common initial energy state $\ket{\psi_0}$, between the ``ideal'' energy state $\ket{\psi_2}$ and the energy state $\ket{\psi_1}$ we can prepare with the physical resources that are available to us (see figure 2). One can prove that $d(U_\beta,U_\alpha) = 2(1-\cos(\beta-\alpha))$. By denoting the error $\epsilon =d(U_\beta,U_\alpha)/2$, it follows that our probability is related to the quantum Born rule: 
\begin{equation}\label{QM2}
 P_{\mathtt{complexity}} = \langle \psi_2 \vert \psi_1\rangle = \sqrt{P_{\mathtt{QM}}}.
\end{equation}

\begin{figure}
\centering
\includegraphics[scale=0.6]{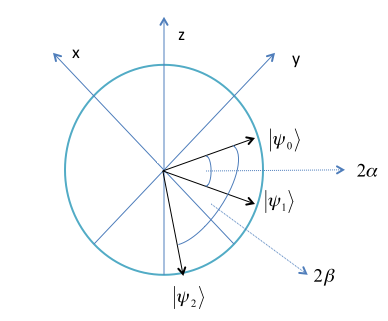}
\caption{Quantum probabilities as distance measures}
\end{figure}

What is the physical meaning of ``preparation'' and ``estimation''? Physically, these are measurements -- in our case, measurements of energy.  But a proper energy measurement probes the time evolution, and therefore cannot be done with arbitrary time, unless one knows {\em in advance} the Hamiltonian of the system at hand.\footnote{Aharonov \& Bohm (1961), for example, proposed a clever way to bypass the time--energy uncertainty principle by measuring a spin--half particle in a magnetic field. Later it was shown that this ability to bypass the time--energy uncertainty relation rests on the prior knowledge of the Hamiltonian: if one knows the Hamiltonian, one need not spend any time (hence computational -- qua physical -- resources) in estimating it, but whenever the Hamiltonian of a system is unknown, determining it to precision $\Delta H$ requires a time $\Delta t$ given by $\Delta t \Delta H \geq 1$. See (Aharonov, Massar, \& Popescu, 2002).} Rather, the time required to perform the measurement and the precision of the measurement are constrained by the time--energy uncertainty relation. For this reason, unless we know {\em in advance} the Hamiltonian that governs the dynamics of a physical system, we can never predict its future behavior with probability $1$. To know the Hamiltonian, however,  we need to measure the system's energy, i.e., to prepare it in a certain energy state, and this preparation will not be error--free, whence objective uncertainty.

In this sense our proposal can serve as a basis for a new interpretation of QM. Recall that on the subjective view of QM, the quantum state is treated as a state of knowledge, and quantum probabilities (calculated by the Born rule) are interpreted as ``gambling bets'' of agents on results of experiments, a-la Ramsey--De Finetti (Fuchs, 2010). In contrast, in Bohmian mechanics, the alternative epistemic approach in the foundations of QM, the probabilities are for particles to have certain positions; they signify our ignorance thereof. The above insistence on ``agents'', or ``observers'' -- considered by proponents of the subjective view as their claim to fame (Fuchs \& Peres, 2000) -- is rejected by Bohmians as ontologically vague; the whole point behind arguments about typicality is to free the discussion from such notions. Our new idea about probability simply avoids this debate altogether by supplying a possible third way: what quantum probabilities are probabilities {\em for} is neither the positions of particles, nor the gambling bets of learned observers. Rather, quantum probabilities simply quantify how hard it is to realize a physical state; they measure the `distance' between the current state of a physical system and any other state thereof, given the resources (energy/time) that are available to that system at that moment. This alternative allows us to interpret quantum probabilities as objective deterministic chances (and in so doing to turn QM once again into a physical theory about the world), without having to support typicality and nonlocal hidden variables.

These two examples demonstrate the promise of our approach. In both cases -- classical and quantum mechanics -- the dynamics is deterministic, and yet probability arises from measurement errors which in turn result from lack of sufficient physical resources. In the classical framework resources are in principle unbounded (potentially infinite) but finite {\em in practice\/}; in the quantum case they are finite {\em in principle\/}.

\subsection{Quantum vs. Classical probabilities}
\noindent In the approach presented here, the origins of both quantum and classical probabilities is identical -- they both stem from objective deterministic chances which supervene on time--complexity classes and relative availability of physical resources. This is not surprising. The physical Church--Turing thesis collapses two separate distinctions, namely, (un)predictability and (in)determinism, into one (Earman, 1986), treating both on a par as epistemic. Some maintain (Pitowsky, 1996) that such an alleged subordination of metaphysics to epistemology makes it is impossible to distinguish between quantum and classical probabilities, and stands in flat contrast to the famous no--hidden--variables results, and the received wisdom about the difference between classical and quantum probabilities, captured by the violations of Bell's inequalities. 

That metaphysics is ignored is a natural consequence of our empiricist framework. But what the critics fail to notice is that an identity in {\em kind} need not entail an identity in {\em measure\/}, and that one can still distinguish between quantum and classical probability measures despite their common origin; metaphysics, we suggest, has nothing to do with this.\footnote{To be fair to Pitowsky (1996), his analysis concerned the notion of computability, while ours concerns the notion of complexity, or efficiency.}  

In the classical case the error (and hence the probability) can decrease (increase) arbitrarily with the increase of these resources. In contrast, in the quantum case the uncertainty principle imposes an actual cut--off on such a potential infinity of resources, which makes it impossible in principle to eliminate the error. Thus the inability of Laplace's demon to predict quantum phenomena results not from some vague metaphysical notion of quantum indefiniteness (``no--hidden--variables"), but from the actual finite bounds on physical resources in our world. Consequently, while the notion of probability here proposed is predicated in both the quantum and the classical cases on error, or lack of sufficient physical resources, this shared origin results not from subordinating ontology to epistemology. Rather, it is the product of a physical cut--off which excludes an unbounded increase of physical resources in the preparation of any given state.
	
That quantum and classical probabilities may be traced to the same origins stems from assumptions the actual bound imposed on physical resources in the world, and the discreteness of energy. In such a finite and discrete world, classical unpredictability that results from ``lack of knowledge" is on a par with quantum uncertainty; both arise from measurement errors as a consequence of insufficient physical resources. But the new view also changes the rules of the game, as it simply rejects vague metaphysical notions such as  ``quantum indefiniteness" as possible criteria for distinguishing quantum and classical probabilities. On the other hand, that a physical system is always in a definite state need not entail hidden variables, as on the view proposed here it is simply not the case that the system is in a definite but {\em unknown} state. Rather, the actual definite state the system possesses is just {\em different} from the ideal state we would like it to be in. It is this discrepancy between the actual and the ideal is what generates probabilities in both the quantum and the classical cases.

Finally, despite the lack of {\em metaphysical} difference, we can still distinguish between quantum and classical probabilities. Instead of ``hidden variables'' vs. ``indefiniteness'' we suggest a difference in {\em measure\/}, which in our case is complexity--induced. Indeed, it is a working hypothesis within quantum information scientists that any classical computation that would be harnessed for the simulation of quantum phenomena would do so inefficiently.\footnote{This conjecture was first voiced by Feynman (1982). Computer scientists have formalized it as $\mathtt{BPP} \subseteq \mathtt{BQP}$ (Aaronson, 2009).} Our approach can easily accommodate such a putative difference: the notion of probability we propose here is {\em defined} as the relative size of the set of time--complexity classes that can realize a physical state. That quantum and classical probabilities share the same origins need not entail that for {\em every} physical state the above relative size is also identical. Quantum dynamical evolutions may ``consume'' (in the sense developed in fn. (\ref{consume})) less resources than classical ones, and so the probability of some physical states may as well be different when realized by quantum or by classical dynamics. We suggest to view the violations of Bell's inequality as designating exactly this difference; a difference in complexity, not in metaphysics (Buhrman {\em et al.\/},  1998). Note, moreover, that such a criterion is completely in accord with our current empirical knowledge, and yet, contrary to its metaphysical counterpart, it leaves open the question of the universality of quantum theory.\footnote{Our probability measure depends on the dimension of the system, which appears to be a key factor in the open problem of scaling--up quantum information processing devices.}

\section{Conclusion: Probability as Distance Measure}
\noindent Our distance measure satisfies Kolmogorov axioms (see appendix B--E), hence, at least mathematically, it is worthy of the name ``probability''. It also explains away ignorance by tying error (in the preparation of the initial state, or propagator) to probability (of the desired state, or propagator), and by supervening this probability on time--complexity and physical resources. In the classical context, it appears to be a natural physical interpretation of the epistemic probabilities that arise in statistical mechanics (see eq. (\ref{ProbRule})). Preliminary results (see eqs. (\ref{QM1}) and (\ref{QM2})) suggest that such a distance measure is also a natural interpretation of quantum probabilities.

We emphasize again that we are only proposing a new {\em interpretation} to the meaning of probability: instead of interpreting probability as a measure of ignorance (which is the standard way in a deterministic dynamical context), we propose an interpretation in terms of the distance (in terms of the relative physical resources) between an actual state and an ideal one. In this sense our proposal is only qualitative. Quantitatively, we {\em conjecture} that one can reproduce the observed relative frequencies and standard results in statistical physics from such a notion which supervenes on time--complexity classes; our two models should be thus seen as plausibility arguments in support of this conjecture.\footnote{Albert (2000, ch. 5) offers a similar conjecture when he proposes that the probabilities of SM supervene on transition probabilities of a more fundamental collapse dynamics. Our view is deterministic, hence excludes collapse, but we too suggest that probabilities in statistical physics are dynamical transition probabilities. In our story, however, they supervene on time--complexity and relative physical resources.}  Moreover, we do not pretend in any way to go beyond objective probabilities in statistical physics in our interpretation. Whatever problems exist in connecting these with the {\em ordinary} notion of probability, namely the connection to relative frequencies, or to betting behavior, also exist in our interpretation, and we do not purport to solve them here.

Concluding, we have argued that he amount of physical resources that separate two physical states is an objective feature of the world, and that computational complexity theory allows us to map this feature onto $[0,1]$. This mapping, as we have shown, has all the characteristics of a discrete probability function, that can be interpreted as a measure of precision and control. 


\section*{Appendix}
\subsection{Measures}
\noindent  Let us consider an arbitrary exponential function $c^n$ and its associated probability:
\begin{equation}
P(c^n)=[\mu-\frac{\arctan (c^n ln c)}{\Pi/2}\mu]\alpha.
\end{equation}
where 
\begin{equation}
\alpha=\frac{1}{\sum_{f_i|f_i \in Poly}\xi_{f_i}+\sum_{f_i|f_i \in Exp}\xi_{f_i}}.
\end{equation}
It is straightforward to show that for a large dimension $n$
\begin{equation}
\forall c: P(c^n) \approx (\mu - \mu)\alpha=0.
\end{equation}
In contrast, since for a large dimension $\frac{\arctan (cn_1^{c-1})}{\Pi/2} \approx 1$, the unnormalized measure $\xi_p$ on  $\mathtt{Poly}$ doesn't change and is $1-\mu_e$, but when we normalize, we get a resolution of the identity
\begin{equation}
P(f(n))=\alpha\xi_p = \frac{1-\mu_e}{i(1-\mu_e)} = i^{-1}
\end{equation}
where $i$ is the number of polynomial functions in $\mathtt{Poly}$.
\subsection{Joint Probability}
\noindent To calculate joint probability in case of independence one needs to realize a new probability space $S_{3}$ from two given probability spaces $S_1$ and $S_2$, where the new couple $\{n_{3}, \left(\frac{E}{t}\right)_{3}\}$ is given by the respective sums, i.e., $n_{3} = n_1 + n_2$ and $\left(\frac{E}{t}\right)_{3} = \left(\frac{E}{t}\right)_1 + \left(\frac{E}{t}\right)_2$.
\begin{itemize}
\item $S_{3}$ is defined accordingly as:
\begin{equation}
S_{\bar n_3} = \{ g \in \Omega | O(g_{\bar n_{3}}) \leq \#(\left(\frac{E}{t}\right)_3)\} 
\end{equation}  
\item The probability measure $P$ is given, again, by the mapping
\begin{equation}
\forall A \in F : P_{\{n_A, \left(\frac{E}{t}\right)_A\}}(A) = \frac{|A|}{|S_3|}
\end{equation}
\item One can calculate the joint probability for  a combined state $C=A\cap B$ by taking into considerations the way the physical resources are distributed between the two components of the combined system. Since we are constructing a new probability space, additional constraints must be satisfied to maintain the appropriate relations between this new space and the earlier, atomic ones.
\end{itemize}
\begin{enumerate}
\item We define equivalent processes to have the same power and the same dimension: 
\begin{itemize} \item $\left(\frac{E}{t}\right)_1 = \left(\frac{E}{t}\right)_2$. 
\item $n_{A\cap B} = n_1 + n_2 = 2n_1=2n_2$.
\end{itemize} 
So in this case, since all interactions are local, the total available power is $\left(\frac{E}{t}\right)_{A\cap B} = \left(\frac{E}{t}\right)_1 \frac{n_1}{n_1+n_2} + \left(\frac{E}{t}\right)_2 \frac{n_2}{n_1+n_2}$. 
\item For such processes, since $n$ increases and the power remains the same, it follows that $P_{3}(A \cap B) < P_1(A)$;  $P_{3}(A \cap B) < P_2(B)$ (where the $P$s are calculated for each case separately).
\begin{figure}
\centering
\includegraphics[scale=0.35]{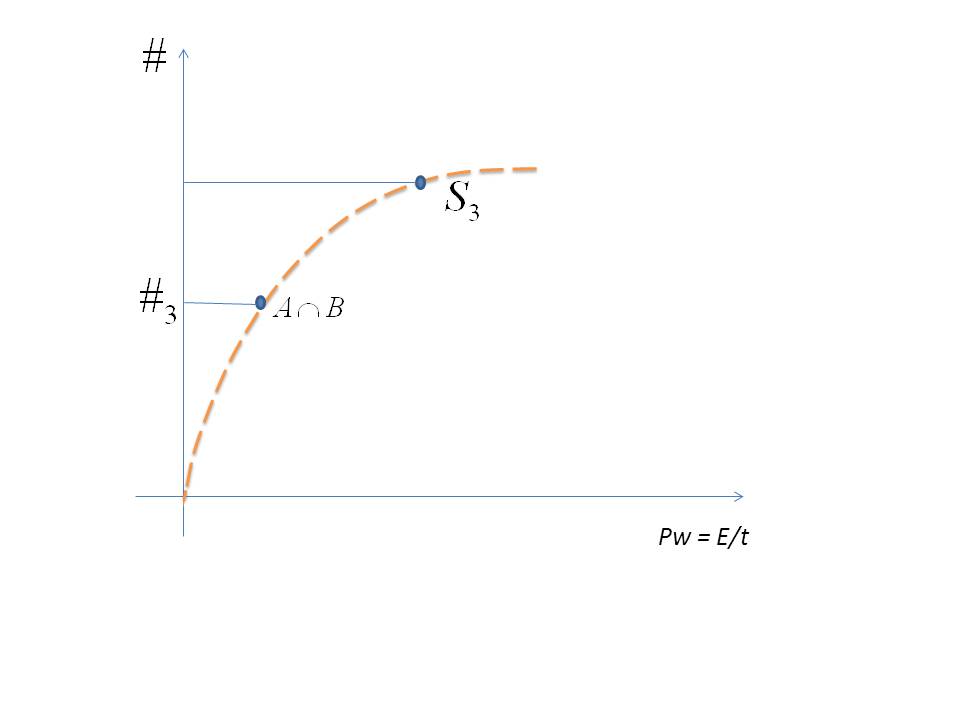}
\caption{Independence}
\end{figure}
\item If $B$ depends on $A$, $P_3(A\cap B) > P_1(A)P_2(B)$. This case is accommodated in our model by noticing that the computation time for $B$ already includes the computation time for $A$, hence the total computation  time is shorter than the computation time in the case of independence, hence the power of the computation is higher than the case of independence, as required.
\end{enumerate}
\subsection{Conditional Probability}
\noindent Conditional probability $P(A|B)$ is calculated by rescaling $S_1$ to fit $S_2$ (in terms of $n$), and by calculating the ratio $P(A \cap B)/P(B)$
\begin{figure}
\centering
\includegraphics[scale=0.35]{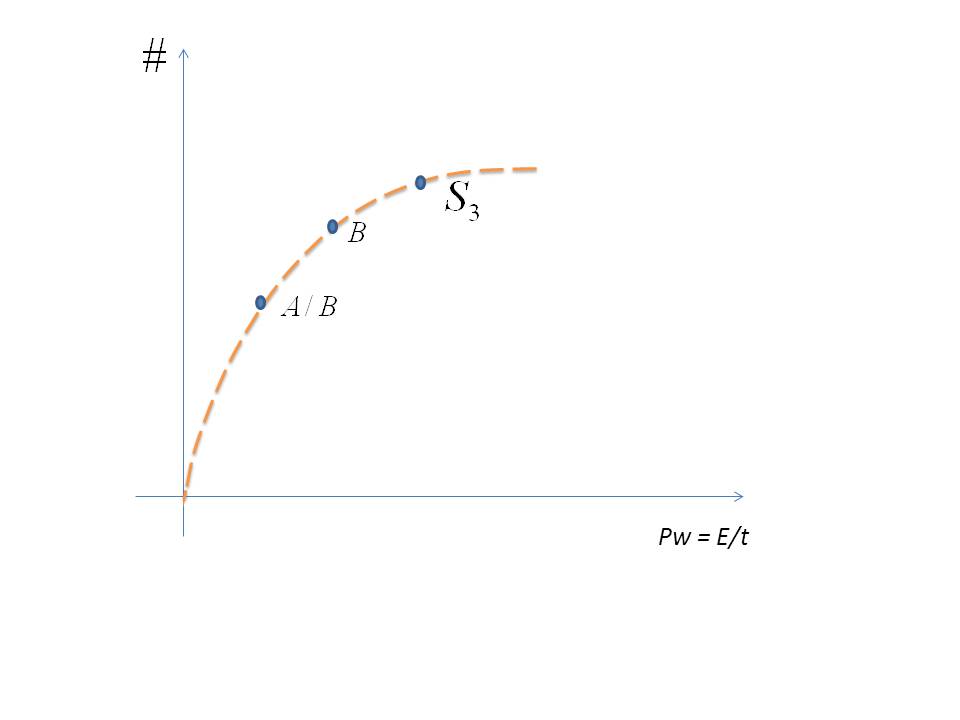}
\caption{Conditional probability}
\end{figure}
\subsection{Constraints}
\noindent These requirements can be used to constrain $f$: when we blow up the dimension by a factor of $\gamma$ and keep the power fixed, the derivative goes down by a factor of $\gamma^{\alpha}$; when we blow up the dimension and keep the derivative fixed, the power goes up by
 a factor of $\gamma^{\frac{\alpha}{\beta-1}}$. 
For non--equivalent processes, where $\left(\frac{E}{t}\right)_A = \delta \left(\frac{E}{t}\right)_B$ ($0<\delta<1$), we can show that 
\begin{equation}
\left(\frac{E}{t}\right)_A < \left(\frac{E}{t}\right)_{A\cap B} <  \left(\frac{E}{t}\right)_B.
\end{equation}
If we require further that the power of $A$ grows {\em less} than the ``blow--up effect'' (that is, that as $n$ increases the Power axis increases {\em less} rapidly than the number--of--steps axis)
\begin{equation}
\frac{\left(\frac{E}{t}\right)_{A\cap B}}{\left(\frac{E}{t}\right)_A}< \left(1 +\frac{1-\delta}{\delta \gamma}\right).
\end{equation}
we get a constraint on $\alpha$ and $\beta$:
\begin{equation}
\frac{\alpha}{\beta-1} > \log_\gamma\left(1+\frac{1-\delta}{\delta \gamma}\right).
\end{equation}

\subsection{Proofs}
\noindent For the sake of simplicity, and without loss of generality, instead of taking the function $\#=(n^\alpha\mathtt{Pw})^{1/\beta}$, we concentrate on the inverse function $\mathtt{Pw} = n^{-\alpha}\#^\beta$, while recalling that for any continuous, derivable, and monotonic function $g$, $(g^{-1})'(x) = \frac{1}{g'(y)}$.

Let $\left(\frac{E}{t}\right)_A = \mathtt{Pw}_A$, $\left(\frac{E}{t}\right)_B = \mathtt{Pw}_B$, and $n_A = \Delta n_B$, where $\Delta\in \mathbb{R}, \Delta>1$ , and let $0<\frac{\mathtt{Pw}_A}{\mathtt{Pw}_B}=\delta\leq1$. We prove that:  $\mathtt{Pw}_A \leq \mathtt{Pw}_{A\cap B} \leq \mathtt{Pw}_B$ (equality holds only when $\mathtt{Pw}_A=\mathtt{Pw}_B$) :
\begin{eqnarray}
\mathtt{Pw}_{A\cap B}=\mathtt{Pw}_A \frac{\Delta n_B}{(\Delta+1)n_B}+\mathtt{Pw}_B\frac{n_B}{(\Delta+1)n_B}=\\ 
=\delta \mathtt{Pw}_B \frac{\Delta}{\Delta+1}+\mathtt{Pw}_B\frac{1}{(\Delta+1)n_B}=\\
=\frac{\Delta\delta+1}{\Delta+1}\mathtt{Pw}_B=\frac{\Delta\delta+1}{\Delta\delta+\delta}\mathtt{Pw}_A.
\end{eqnarray}
But,  
\begin{equation}
\frac{\Delta\delta+1}{\Delta+1}\leq 1\;\;\;  ;  \;\;\; \frac{\Delta\delta+1}{\Delta\delta+\delta}\geq 1.
\end{equation}

If we blow up $n$ by a real factor $\gamma>1$  and we keep the power fixed:
\begin{equation}
f_{\bar n}(\#)=\frac{1}{\bar n^\alpha} \#^\beta \;\;\; ; \;\;\; f^\prime_{\bar n}(\#)=\frac{\beta}{\bar n^\alpha}\#^{\beta-1}.
\end{equation}
Analogously:
\begin{equation}
f^{\prime}_{\gamma \bar n}(\#)=\frac{\beta}{{(\gamma \bar n)}^\alpha}\#^{\beta-1}.
\end{equation}
So, 
\begin{equation}
\frac{f^{\prime}_{\gamma \bar n}(\#)}{f^{\prime}_{\bar n}(\#)}=\gamma^{-\alpha}.
\end{equation}
If we blow up $n$ by a real factor $\gamma>1$  and we keep the derivative fixed:
\begin{equation}
\mathtt{Pw}=f_{\bar n}(\#)=\frac{1}{\bar {n}^\alpha}\#^\beta.
\end{equation}
The derivative for a number of steps $\bar \#$ will be 
\begin{equation}
f^{\prime}_{\bar {n}}(\bar \#)=\frac{\beta}{\bar n^\alpha}\bar \#^{\beta-1}.
\end{equation}
Analogously, for the new (blown--up) power $\tilde{\mathtt{Pw}}$
\begin{equation}
\tilde{\mathtt{Pw}}=f^{\prime}_{{\gamma \bar n}}(\tilde \#)=\frac{\beta}{{(\gamma \bar n)}^\alpha}\tilde \#^{\beta-1}.
\end{equation}
In order to maintain the same derivative, we find the value of $\tilde \#$ such that:
\begin{equation}
\frac{\beta}{{(\gamma \bar n})^\alpha}\tilde \#^{\beta-1}=\frac{\beta}{\bar n^\alpha}\bar \#^{\beta-1}. 
\end{equation}
We obtain: $\tilde \#=\gamma^{\frac{\alpha}{\beta-1}}\bar \#$
So, 
\begin{equation}
f_{\gamma \bar n}(\gamma^{\frac{\alpha}{\beta-1}}\bar \#)=\frac{1}{(\gamma \bar n)^\alpha}(\gamma ^{\frac{\alpha}{\beta-1}}\bar \#)^\beta= \gamma^{\frac{\alpha\beta}{\beta-1}}f_{\bar n}(\bar \#).
\end{equation}
Hence, $\frac{\tilde{\mathtt{Pw}}}{\mathtt{Pw}}=\gamma^{\frac{\alpha}{\beta-1}}$.
Finally, we know that $\frac{\mathtt{Pw}_{A\cap B}}{\mathtt{Pw}_A}=\frac{\Delta\delta+1}{(\Delta+1)\delta}$ and that $\frac{\tilde{\mathtt{Pw}_A}}{\mathtt{Pw}_A}=(\Delta+1)^{\frac{\alpha}{\beta -1}}$. Let us recall that $\gamma=\Delta+1$ and that $0<\delta \leq 1$.
We have to prove that there is some constraint on $\alpha$ and $\beta$ such that $\frac{\mathtt{Pw}_{A\cap B}}{\mathtt{Pw}_A} < \frac{\tilde{\mathtt{Pw}_A}}{\mathtt{Pw}_A}$.
From 
\begin{equation}
\frac{a\delta+1}{(\Delta+1)\delta} < (\Delta+1)^{\frac{\alpha}{\beta -1}}.
\end{equation}
we obtain that
\begin{equation}
\frac{\alpha}{\beta-1} > \log_{\gamma}\frac{\Delta\delta+1}{\Delta\delta+\delta}.
\end{equation}
But 
\begin{equation}
\frac{\Delta\delta+1}{\Delta\delta+\delta}=\frac{\Delta\delta+1+\delta-\delta}{\delta(\Delta+1)}=\frac{\delta(\Delta+1)+1-\delta}{\delta(\Delta+1)}=1+\frac{1-\delta}{\delta \gamma}. 
\end{equation}
so our constraint is 
\begin{equation}
\frac{\alpha}{\beta-1} > \log_\gamma\left(1+\frac{1-\delta}{\delta \gamma}\right).
\end{equation}


\end{document}